# Metasurface-dressed two-dimensional on-chip waveguide for free-space light field manipulation

Yimin Ding,[†‡] Xi Chen,[†‡] Yao Duan,[†‡] Haiyang Huang,[†‡] Lidan Zhang,[†] Shengyuan Chang,[†] Xuexue Guo,[†] and Xingjie Ni[†ƎÚ]

[†] Department of Electrical Engineering, The Pennsylvania State University, University Park, PA 16802, USA.

[Ǝ] Material Research Institute, Pennsylvania State University, University Park, PA 16802, USA.

[‡] These authors contributed equally.

[Ú] *Corresponding author:* xingjie@psu.edu

ABSTRACT

We show that a metasurface-coated two-dimensional (2D) slab waveguide enables the generation of arbitrary complex light fields by combining the extreme versatility and freedom on wavefront control of optical metasurfaces with the compactness of photonic integrated circuits. We demonstrated off-chip 2D focusing and holographic projection with our metasurface-dressed photonic integrated devices. This technology holds the potential for many other optical applications requiring 2D light field manipulation with full on-chip integration, such as solid-state LiDAR and near-eye AR/VR displays.

KEYWORDS

Metasurfaces, photonic integrated circuits, two-dimensional waveguides, meta-lenses, meta-holograms.

Metasurfaces[1,2] composed of artificially engineered light-emitting pixels arbitrarily programmed provide versatile platforms for both fundamental research and practical applications. The metasurface allows complete control over electromagnetic field properties, including phase, amplitude, and polarization. It enables various applications such as metalenses[3,4], holograms[5-7], OAM generations[1,8], and polarimeters[9,10]. The functional metasurface-based flat optics can be made ultrathin, but most demonstrated ones cannot work without free space[11], making it difficult for fully on-chip integration.

The photonic integrated circuits (PICs)[12,13], where light is wired on a chip through functional optical components such as waveguides, modulators[14], and multiplexers/demultiplexers[15], provide a long-term solution for fast-growing demands on bandwidth and speed in information systems in recent years. PICs could potentially reduce the complexity and cost, and increase the stability, repeatability, and scalability for future high-yield production. Building interfaces between free-space light and guided modes propagating on chips is crucial to realize PIC-based devices with extended functionality. Currently, mature coupling technologies between PIC and free space are mainly based on edge coupling[16] and grating coupling[17], whose functions are limited due to the lack of complete control on scattered electromagnetic waves.

By combining the versatility of metasurfaces and the compactness of PIC-based systems, new platforms with improved functionalities have been demonstrated. For example, an in-plane mode converter based on metasurface coated waveguides has been shown[18]. Similarly, researchers also have shown a compact mode-division multiplexing PIC and on-chip wavefront shaping components[19-21]. We have recently demonstrated a versatile platform where free-space light can be flexibly molded into arbitrary free-space optical modes by exploiting the collective behavior of nanoantennas coated on top of a waveguide[22]. We showed flexible free-space light manipulation, including directional coupling, off-chip focusing, and optical vortex generation, from guided waves in a metasurface-coated single-mode ridge waveguide. In contrast to the in-plane meta-waveguides that could only manipulate waves without involving free space, nanoantennas integrated onto waveguides based on their particular property.

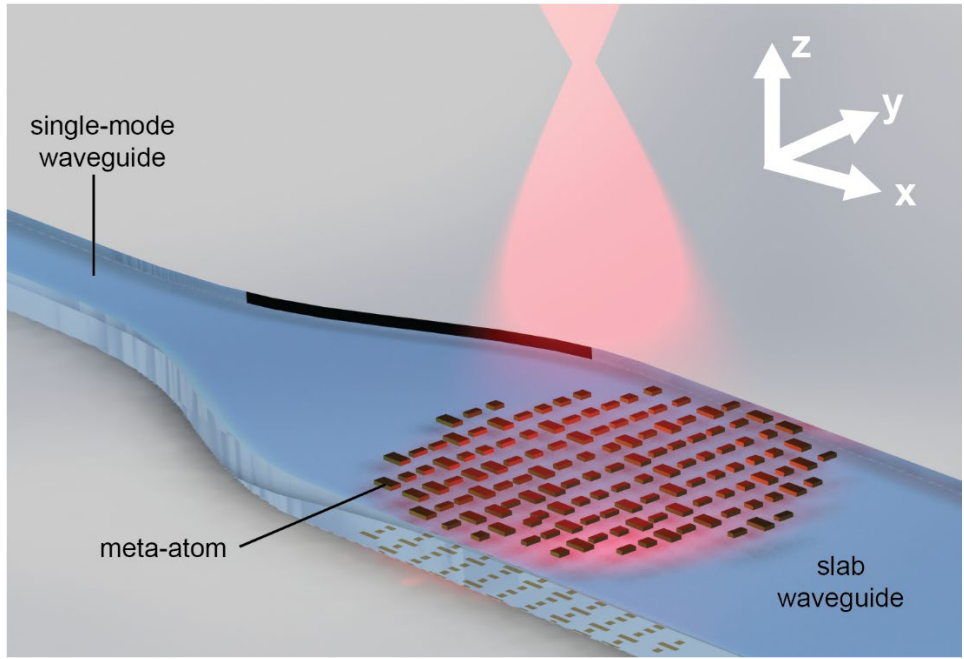

Figure 1. Schematic plot of 2D guided-wave-fed metasurface. Propagating wave injected into the slab waveguide interacts with the metasurface on top are scattered with designed wavefront for a specific function (off-chip focusing here).

Our earlier work provided a practical approach to realize one-dimensional (1D) light field phase manipulation. Furthermore, polarization and wavelength (de)multiplexing between free space and one-dimensional ridge waveguide were also reported[23-25]. However, light field control through two dimensions with higher flexibility is required for further advanced applications. For instance, light detection and ranging (LiDAR)[26,27] with a long propagation distance are more in favor of a larger device aperture size in both dimensions to control the diffraction. Near-eye holographic display for augmented reality (AR) / virtual reality (VR)[28,29] also requires two-dimensional (2D) phase control. Although constructing arrayed single-mode waveguides could be a potential approach to realize 2D wave manipulation, different phase offsets and neighboring couplings between adjacent waveguides need to be addressed by phase compensation such as electro-photo modulation, which increases system complexity and hinders actual pixel density. In this work, we extended the capability of using metasurface to control scattered light from 1D to 2D. By placing metal-dielectric-metal meta-atoms on top of a slab waveguide, we can tune the phase of scattered light covering $2\pi$. Compared to our earlier work, the mode conversion and intensity distribution inside the waveguide were investigated for 2D manipulation. Based on these, we demonstrated different functions such as 2D metalenses, as conceptually shown in Figure 1, and holograms on waveguides experimentally or numerically. Our work showed that with 2D light field manipulation capability, the proposed structure was able to realize more functions.

Starting from a simple case, we considered a guided wave of fundamental mode propagating inside a slab waveguide having a plane wavefront perpendicular to the propagation direction $x$, with a propagation constant $\beta$. We then put a 2D metasurface on top of the slab waveguide, and the guided wave in the waveguide evanescently coupled with the nanoantennas of the metasurface. Being excited, the nanoantennas extracted a small portion of the guided light into free space with imposed phase lags. Therefore, we were able to construct a metasurface composed of nanoantennas that could locally tune the phase $\varphi_{an}(x, y)$ as

$$\varphi_{an}(x, y) = \phi(x, y) - \beta x = \phi(x, y) - n_{mode} \frac{2\pi}{\lambda} x \qquad (1)$$

where $\phi(x, y)$ is desired phase profile for realizing specific function, $n_{mode}$ is the effective mode index of this fundamental mode, and $\lambda$ denotes the free-space wavelength. The nanoantennas we used here are metal-insulator-metal (MIM) sandwiched nanobars, supporting both electric and magnetic dipolar resonances around the desired operational wavelength (1550 nm)[22]. By changing the dimensions (length and width) of the nanobars, we get a phase lag $\varphi_{an}(x,y)$ from each nanoantenna that can range from 0 to $2\pi$. To design the device, we built a phase-amplitude response library with regard to the nanoantenna dimensions using full-wave simulations with a commercially available finite element method (FEM) solver COMSOL Multiphysics. A sandwiched nanobar composed of stacked Au, Si, and Au layers, with a thickness of 30 nm for each layer, was placed on top of a 500-nm-thick Si slab waveguide on SiO$_2$ substrate. In our simulation, a fundamental TE mode, numerically obtained by an eigenmode solver, was excited at the input port of the Si slab waveguide and propagated through the waveguide. The modal index of this TE$_{00}$ mode $n_{mode}$ was also provided by the solver. The scattered light from the nanobar was monitored right above the center of the nanobar. We obtained both the amplitude and phase of the scattered electric field. Keeping the thickness of the nanobar unchanged, we swept the width (90 nm $\leq l_x \leq$ 130 nm) and length ($100 nm \leq l_y \leq 400 nm$) of the nanobar to get a phase/amplitude-geometries map. Taking into account the nanofabrication imperfection, trapezoidal vertical cross-sessions and fillets of the nanobar were built in our numerical model. Based on the phase/amplitude-geometries map, three different nanobars geometries with the same scattered electric field amplitude and phases equaling $2\pi/3$, 0, and $-2\pi/3$, respectively, were selected as the building blocks for the metasurface. Therefore, an arbitrary phase profile expanding through the 2D functional area could be realized by spatially arranging this set of three nanoantennas with different geometrical parameters. It is worth pointing out that to achieve the precise control of the desired phase profile, the field distribution of the input guided wave needs to be taken into consideration throughout the region. For simplicity, we kept the wavefront inside the slab waveguide flat through the propagation in our proof-of-concept demonstration. To assure this, (1) the input mode needs to be purely fundamental without any other modes, and (2) the metasurface on top of the waveguide does not affect the propagating mode remarkably. In other words, the influence of the nanobars on the guided wave is negligible to avoid mode conversion inside the slab waveguide. Otherwise, the phase provided by the propagation $\varphi(x, y) = \beta x$ could be inaccurate.

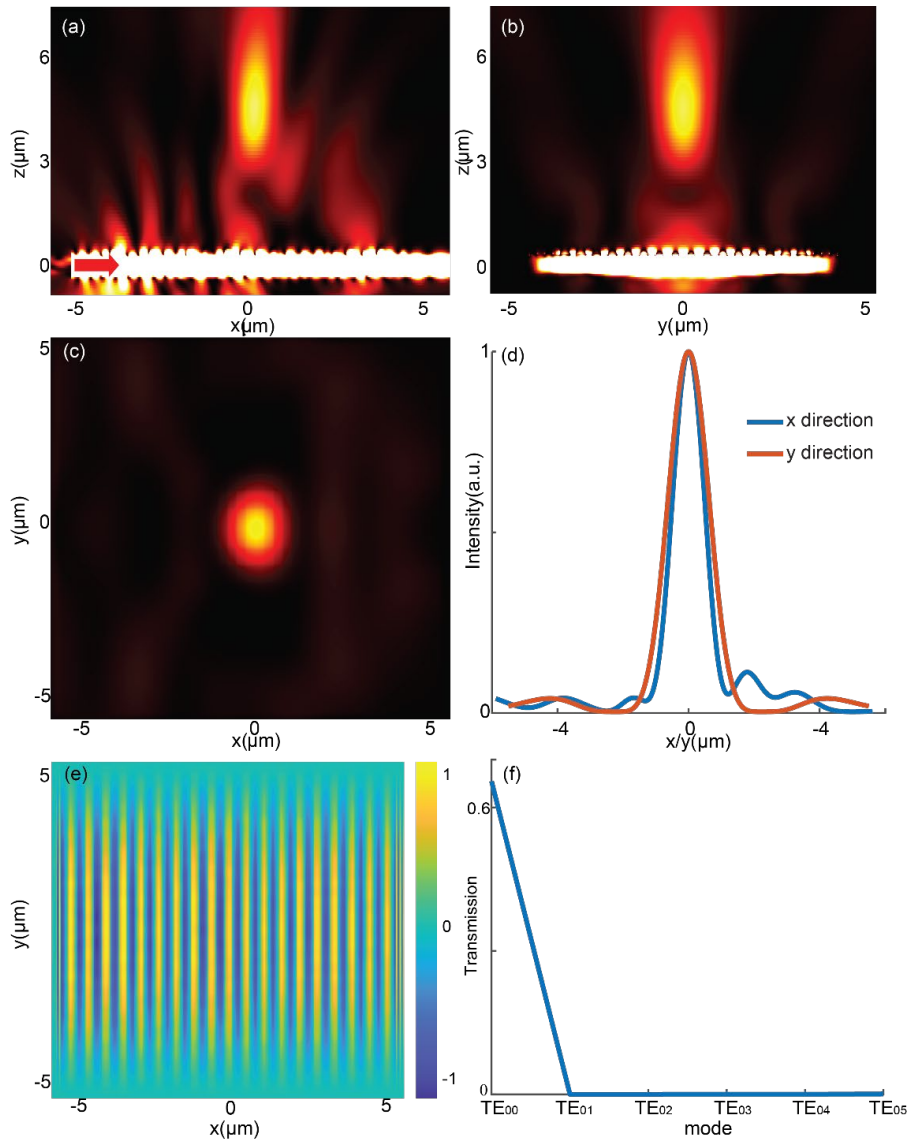

Figure 2. Numerical simulation of 2D guided-wave-fed metalens. The light propagates along the x-axis from left to right as the red arrow in (a). Scattered light intensity distribution at a cross-

section of X-Z plane(a), Y-Z plane(b), and focal plane X-Y plane(c) shows an off-chip focused light beam scattered out from the bottom slab waveguide. (d) Intensity distribution along the x and y direction at the focal plane shows a tightly focused airy disk with a radius of about 2.1 μm. (e) The real part of the guided wave electric field underneath the metasurface region shows no evident distortion. (f) Transmission intensity for different modes after the metasurface region. Only fundamental $TE_{00}$ mode was detected at the end of the waveguide.

As a proof-of-concept demonstration of our proposed schematics on 2D on-chip light field manipulation, we first numerically and experimentally demonstrated an integrated 2D metalens on a waveguide. Then, we presented a holographic projection directly from a slab waveguide for a more general manifestation.

To generate a focused beam in free space, the phase of light scattered from the metasurface has a distribution described below

$$\phi(x,y) = -\frac{2\pi}{\lambda}\sqrt{x^2 + y^2 + f^2} \qquad (2)$$

where $f$ is the distance between the device plane and the focal plane, i.e., focal length. We grided our device plane into subwavelength pixels, in which the phase delay in each pixel $\varphi_{an}(x,y)$ can be calculated following equations (1) and (2). We rounded the required phase $\varphi_{an}(x,y)$ to the nearest discrete phase levels ($2\pi/3$, $0$, $-2\pi/3$) and chose the nanobar geometries accordingly. Then the entire device was established by arranging this set of nanoantennas on top of a 500-nm-thick silicon slab waveguide on a glass substrate. As a proof of concept, we numerically simulated a 5 μm × 5 μm 2D metalens on a silicon waveguide with a focal length of 5 μm. The period along and perpendicular to the propagation direction is 220 nm and 440 nm, respectively. $TE_{00}$ mode light injected into the slab waveguide from its left port propagated through the metasurface region.

The simulation results show that a portion of the guided wave was extracted directly by the metasurface on top of the slab waveguide. The extracted light was focused on a focal spot at ~5 μm above the slab waveguide (Figure 2(a)-(c)). The intensity distribution at the focal plane manifested an Airy disk with a full-width-half-maximum (FWHM) of about 2.1 μm, which matched our theoretical prediction well (Figure 2(d)). We conducted mode decomposition for the guided wave in the slab waveguide after the metasurface. The result in Figure 2(f) shows that the transmitted field barely had high-order modes (transmitted intensity < 0.01%) other than fundamental $TE_{00}$ (transmitted intensity > 67%). The electric field, of which the real part is shown in Figure 2(e), underneath the metasurface did not show noticeable distortion, manifesting that the nanoantennas we used here induce negligible mode conversion inside the waveguide.

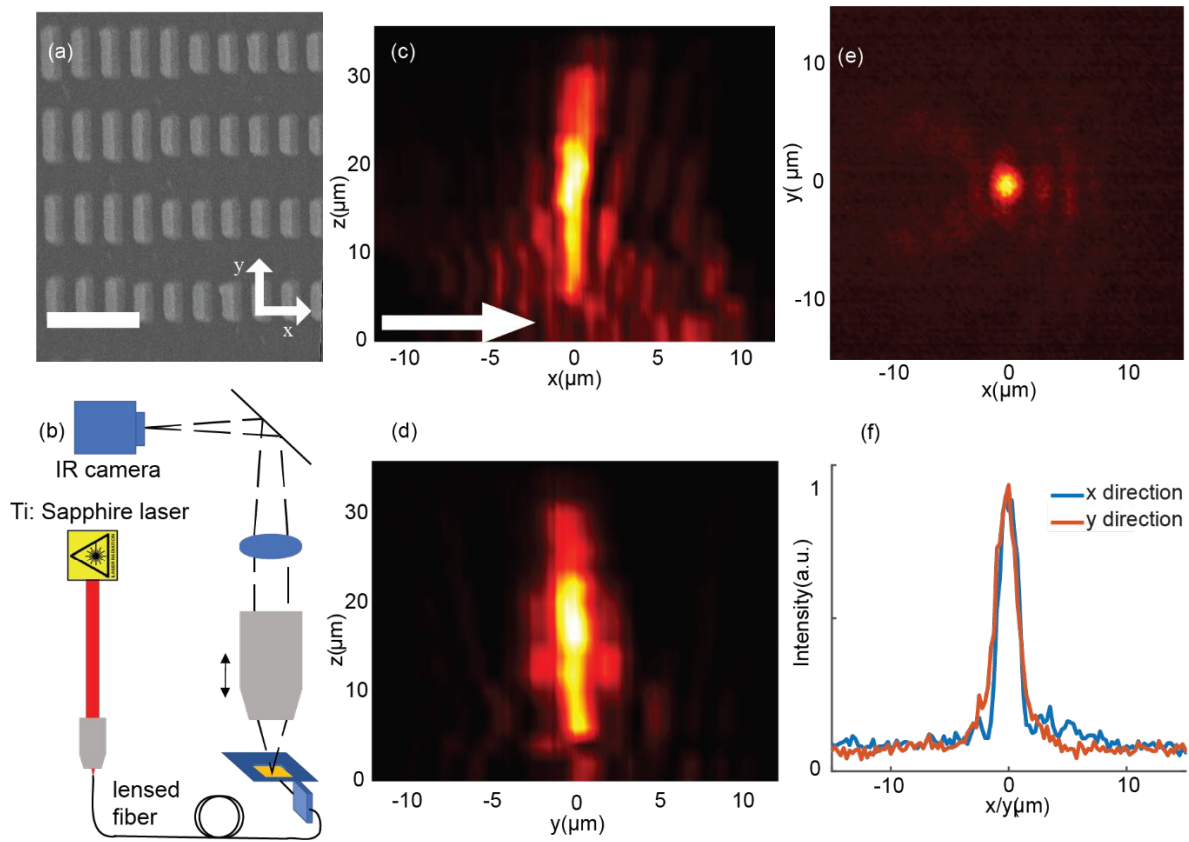

Figure 3. Experimental demonstration of guided-wave-fed 2D metalens. (a) SEM image of the metasurface. The scale bar is 500nm. (b) Optical setup containing a movable objective for reconstruction 3D intensity distribution on top of the device. Measured intensity distribution at the cross-section of X-Z plane(c), Y-Z plane(d), and focal plane X-Y plane(e). (f) Intensity distribution along x and y direction at the focal plane shows a focal spot with full width at half maximum of about 1.8 μm.

We designed and fabricated a guided-wave-fed 2D metalens using two electron beam lithography steps with precise alignment to define the silicon waveguide and the metasurface with processes similar to those in Ref. 22. The SEM image of fabricated metasurface on the waveguide is shown in Figure 3(a). The metasurface region has a size of 17.6 μm × 17.6 μm, and the focal length was designed to be 20 μm. An edge-coupling port connected with a single-mode waveguide was used to eliminate higher-order propagation modes. The single-mode waveguide was immediately followed by a taper which could adiabatically convert the fundamental $TE_{00}$ mode from the single-mode waveguide into a $TE_{00}$ mode with extended width in the slab waveguide[30]. Both two parts guaranteed that the guided wave impinging the metasurface region had a flat wavefront. We obtained three-dimensional (3D) scattered-out light field distribution above the device using a home-built optical setup, as shown in Figure 3(b). A free-space laser beam of 1550 nm emitted from a Ti:Sapphire laser-pumped optical parametric oscillator (OPO) was coupled into a commercially available polarization-maintaining taper-lensed single-mode fiber. The focused laser beam from the tapered end of the fiber was edge-coupled into the input port of our fabricated ridge waveguide sample in an end-fire approach using a 3D translational stage. The light scattered into free space by metasurface on the slab waveguide was collected by an objective (N.A. = 0.95) and then imaged by a tube lens onto an infrared camera. The objective was mounted on a translational stage with a high-resolution piezo-controlled actuator in the *z*-direction. By moving the z stage, 2D real-space images were taken at different heights, and a 3D volumetric image was reconstructed by stacking the 2D images. After that, we extracted the intensity distribution at X-Z and Y-Z cross-section, as shown in Figures 3 (c) and (d), respectively. Our experimental result showed a similar intensity distribution profile as that in our simulations (Figure 2). We observed the light scattered out by the metasurface gradually focused on the height around 18 μm, as shown in Figure 3(e), close to the designed focal length. The FWHM of the focal spot intensity distribution at the focal plane is about 1.8 μm as shown in Figure 3(f), which is slightly larger than the theoretical expectation (1.36 μm).

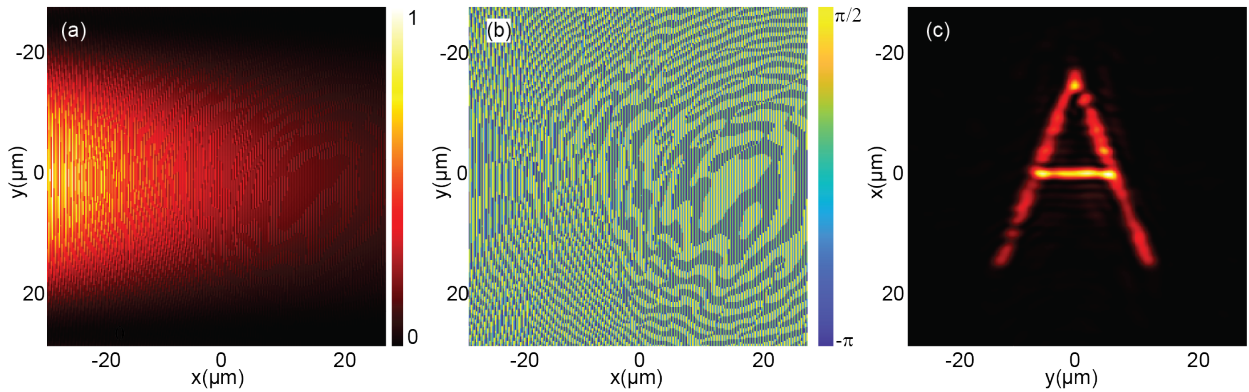

Figure 4. Holographic generation by 2D guided-wave-fed metasurface. The amplitude(a) and phase(a) distribution of the electric field distribution 100 nm above metasurface decorated waveguide. (c) holographic image reconstructed at 20 μm above the metasurface decorated waveguide.

More generally, by leveraging the freedom to control the phase profile of scattered light, we demonstrated a 3D holographic projection directly from a guided-wave-fed metasurface. Hologram-based near-eye display generating 3D images carrying depth information is a promising solution for the vergence-accommodation-conflict (VAC) in AR/VR[31], and incorporating hologram in the form of compactness and lightweight is crucial for developing next-generation head-mounted-devices (HMD). By mapping the phase profile of a holographic image onto the metasurface-coated waveguide, we demonstrated the capability of utilizing our platform for this specific application. First, we calculated the intensity distribution in the slab waveguide. In our slab waveguide configuration, the intensity distribution was a decaying Gaussian function, which can be approximated represented by

$$I(x,y) = Ae^{-\frac{y^2}{\sigma}-\alpha x} \qquad (3)$$

The term $Ae^{-\frac{y^2}{\sigma}}$ denotes a 1D Gaussian function along the *y*-axis centered at y = 0, where the constants $A$ and $\sigma$ were numerically calculated based on a TE$_{00}$ mode profile. The term $e^{-\alpha x}$ here describes the energy decaying through propagation, in which $\alpha$ is the attenuation coefficient estimated by averaging scattering and ohmic loss from different nanoantennas. Then, we calculated the required phase distribution at the metasurface plane using a modified Gerchberg–Saxton (G-S) algorithm for a required image floating at a specific height. In this step, the intensity distribution in the slab waveguide served as the source intensity. The goal was to get the source phase. The intensity and phase formed a complex-valued wavefront which generates desired intensity distribution (target image) at the image plane after diffraction. Then, the final source phase profile was calculated by calculating forward and backward propagation between the source plane and target plane with fixed intensity distribution iteratively. The diffraction propagation between the source plane and the image plane was modeled using the angular spectrum method. The source phase profile converged to the results after several iterations. The phase needs to be compensated from the propagation phase inside the slab waveguide was calculated according to equation (1). After that, the geometric parameters of the antennas in each 220 nm × 440 nm unit cell were determined. We constructed a 28.16 μm × 28.16 μm sized hologram, which corresponding to 128 × 64 pixels, on a slab waveguide. In our hologram calculation using Matlab, to get a more accurate phase profile representation, we used four different nanoantennas with a phase lag of $-\pi, -\pi/2, 0, \pi/2$. It is worth mentioning that, in the hologram construction, the four-level nanoantenna set did not have a constant scattering efficiency. Despite that the amplitude distribution was quite nonuniform (Figure 4(a)) and the phase distribution was discretized (Figure 4(b), we got a clear holographic imaging floating 20 μm above the metasurface plane (Figure 4(c)).

The 2D light field manipulation enabled by combining metasurfaces and waveguides with flexible phase controllability and functionality was demonstrated. This technique allows free-space 2D structured light to generate directly from an integrated photonic chip within a compact form factor. To assure accurate phase manipulation, we kept the metasurface perturbation to the guided wave small enough to avoid mode conversion in this proof-of-concept demonstration. However, this is not a stringent requirement. For instance, more significant perturbation was also considered in the design process, using the electromagnetic inverse design method[32]. Furthermore, due to the energy decaying inside the waveguide, the actual device size is inversely proportional to the scattering efficiency of the metasurface. It is worth noting that as our system is linear and time-invariant, possessing time-reversal symmetry, it is reciprocal and able to convert free-space modes into waveguide modes. In fact, free-space modes to guided modes conversion using 1D metasurfaces on a waveguide has been demonstrated numerically[23].

In conclusion, we demonstrated functional 2D metasurfaces integrated on a slab waveguide driven by a guided wave. An off-chip focusing function was demonstrated numerically and experimentally. The designed numerical aperture is 0.4, and the focused beam has a near-diffraction-limited spot size in the experiment. For a more general phase control demonstration, we also numerically constructed holographic projection using this approach. The calculation shows that this platform can be used for hologram generation with a decaying energy distribution and discretized (4-level) phase profile. Therefore, the developed platform could be exploited for many other optical applications requiring 2D light field manipulation with full on-chip integration, such as solid-state LiDAR and head-mounted AR/VR displays. Furthermore, it is possible to incorporate detour phase metasurfaces[33] as well as tunable metasurfaces[34,35] on a waveguide, which could lead to more degrees of freedom and new applications.

ACKNOWLEDGMENT

The work was partially supported by the Moore Inventor Fellow award from the Gordon and Betty Moore Foundation, the National Aeronautics and Space Administration Early Career Faculty Award (NASA ECF) under grant no. 80NSSC17K0528, the Office of Naval Research (ONR) Basic Research Challenge (BRC) under grant no. N00014-18-1-2371, and the National Institute On Aging of the National Institutes of Health under R56AG062208.